\documentclass[pdflatex,iicol,sn-basic]{sn-jnl}

\usepackage{graphicx}%
\usepackage{multirow}%
\usepackage{amsmath,amssymb,amsfonts}%
\usepackage{amsthm}%
\usepackage[title]{appendix}%
\usepackage{xcolor}%
\usepackage{textcomp}%
\usepackage{manyfoot}%
\usepackage{booktabs}%
\usepackage{algorithm}%
\usepackage{algorithmicx}%
\usepackage{algpseudocode}%
\usepackage{listings}%
\usepackage[caption=false]{subfig}%
\usepackage{pifont}%
\usepackage{tabularx}%
\usepackage{multirow}%
\usepackage{placeins}%
\usepackage{soul}
\usepackage{etoolbox}

\makeatletter
\let\orcidlogo\@undefined

\pretocmd{\@maketitle}{%
  \begin{flushright}
    \normalfont DESY-26-092
  \end{flushright}
  \vspace{-0.3cm}%
}{}{}

\makeatother

\usepackage{orcidlink}

\newcommand{\code}[1]{{\ttfamily\upshape #1}}

\begin{document}

\title{Lossless Compression Performance for PETRA III Datasets}

\author*[1]{\fnm{Malte} \sur{Buschmann}\orcidlink{0000-0002-2036-4763}}
\email{malte.buschmann@desy.de}
\author[1]{\fnm{Yannis} \sur{Schumann}\orcidlink{0000-0002-2379-200X}}
\author[1]{\fnm{Christian} \sur{Voss}}
\email{christian.voss@desy.de}
\author[1]{\fnm{Tigran} \sur{Mkrtchyan}\orcidlink{0000-0002-6217-8539}}
\email{tigran.mkrtchyan@desy.de}
\author[1,2]{\fnm{Philipp} \sur{Neumann}\orcidlink{0000-0001-8604-8846}}
\email{philipp.neumann@desy.de}

\affil[1]{\orgname{Deutsches Elektronen-Synchrotron DESY}, \orgaddress{\city{Hamburg}, \country{Germany}}}
\affil[2]{\orgdiv{High Performance Computing \& Data Science}, \orgname{University of Hamburg}, \orgaddress{\country{Germany}}}

\abstract{Large-scale research facilities increasingly face the challenge of managing rapidly growing data volumes while maintaining sustainable archival infrastructures. We present the first comprehensive study of data heterogeneity and lossless general-purpose compression performance for representative datasets from the PETRA III synchrotron radiation source. Our corpus comprises more than 212~TiB of raw and processed data from ten experiments spanning multiple beamlines, detector systems, and scientific workflows.
We observe substantial heterogeneity both between and within experiments, resulting in compression ratios that vary by more than two orders of magnitude across datasets. Evaluating nine widely used lossless compression tools, we find that Zstandard and LZ4 consistently occupy the high-throughput region of the Pareto front, whereas ZPAQ achieves the highest compression ratios. Furthermore, heterogeneous compression strategies that adapt compressor choice to the underlying file category outperform uniform compression policies.
Extrapolating from the benchmarked datasets to the full PETRA III non-tape storage system, we estimate achievable compression ratios ranging from approximately 1.6 at $\sim$900 MiB/s throughput to 2.1 at $\sim$2 MiB/s. These results provide a quantitative basis for future archival and storage decisions at PETRA~III, its future successor, PETRA~IV, and other large-scale scientific facilities.}


\maketitle

\section*{Statements and Declarations}
\bmhead{Competing interests}
The authors declare no conflicting interests.

\section*{Acknowledgements}
The authors would like to thank the following people for their generous support with data and advice during this project (ordered alphabetically by first name):  Andre Rothkirch, Fabian Westermeier, Fabian Wilde, Gerhard Morozov, Jürgen Hannapel, Linus Pithan, Mwai Karimi, Olof Gutowski, Philipp Middendorf, Pontus Fischer, Tim Salditt, and Tim Schoof.

%%%%%%%%%%%%%%%%%%%%%%%%%%%%%%%%%%%%%%%%%%%%%%%%%%%%%%%%%%%%%%%%%%%%%%%%%%%%%%%%%%%%%%%%%%%%%%%%%%%%%%%%%%%%%%%%%%%%%%%%%%%%%%%%%%%%%%%%%%%%%%%%%%%%%%%%%%%%%%%%%%%%%%%%%%%%%%%%%%%%%%%%%%%%%%%%%%%%%%%%%%%%%%%
\section{Introduction}
\label{sec:introduction}
Synchrotron radiation has enabled major advances across a broad range of scientific disciplines, including medicine \citep{Guenther2021, Prester2024}, engineering \citep{Dallari2023, Martinelli2023}, and materials science \citep{Boudinot2022, Museur2024}.
Large-scale facilities, such as PETRA III in Germany \citep{Franz2006} or the ESRF in France \citep{Cloetens2025}, provide high-brilliance synchrotron radiation to users from different research domains and therefore leverage synergies between the respective research communities.

\begin{figure*}[t]
    \centering
    \includegraphics[width=0.9\textwidth]{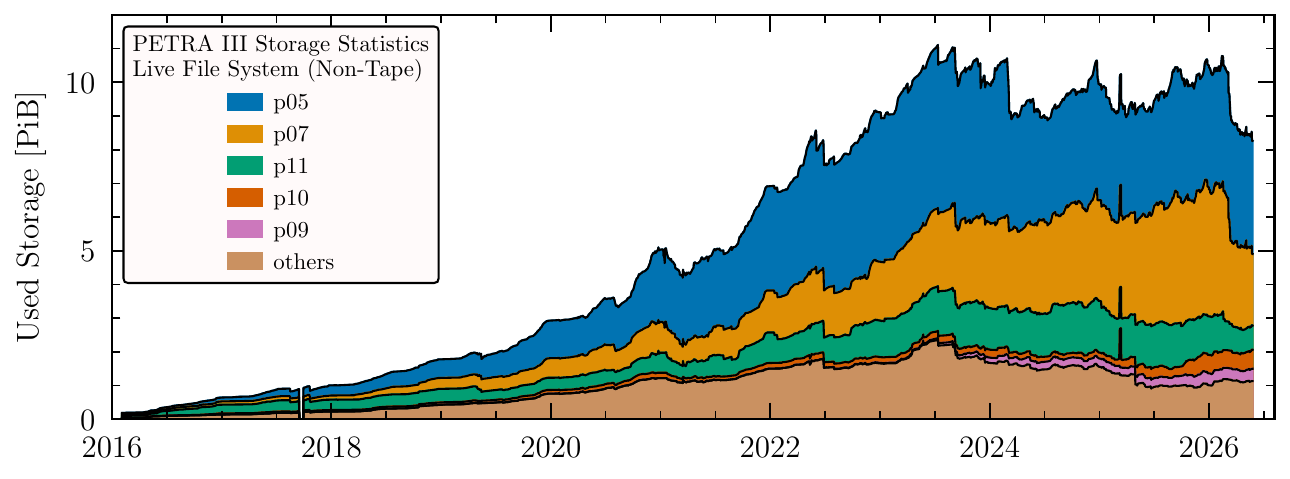}
    \caption{Amount of storage required to store PETRA III data (\emph{raw} + \emph{processed}) at the DESY computing center (online filesystem only, tape data excluded). Colors indicate different experimental endstations, with those included in this study highlighted.}
    \label{fig:asap3_statistics}
\end{figure*}

In recent years, such facilities have served a large number of users (e.g., $6,500$ user visits between Mar 15 and Dec 22 for PETRA III in 2021) while continuously integrating further technical improvements \citep{Schroer2018}. 
The amount of data generated by synchrotron experiments and subsequent analyses has increased continuously over the past years. Maintaining long-term access to these datasets in accordance with the FAIR data principles \citep{Wilkinson2016} therefore contributes significantly to the operational costs of facility computing centers.
At PETRA III, the DESY computing center temporarily stores data on a GPFS-based online filesystem for 180 days after acquisition to enable timely analysis, after which the data are archived to tape with a minimum retention period of ten years. In the past, the volume of data stored on the GPFS filesystem approximately doubled every two years, see Fig.~\ref{fig:asap3_statistics}.

To mitigate the escalating storage demands and the associated costs, the application of data compression techniques or optimized data layouts is essential. 
At ESRF, the BLISS framework \citep{Guijarro2018} provides an integrated solution for data acquisition and export. At DESY, however, the broad heterogeneity of experiments and user-specific software requirements has complicated the adoption of similarly standardized data workflows and storage formats, including optimized container formats such as HDF5 \citep{hdf5}.
As a prerequisite for the future establishment of a unified data solution at PETRA III and PETRA IV, we collected a corpus of representative data from all major data-generating experimental endstations and
\begin{itemize}
    \item conducted a comprehensive survey of data layouts and file types across experimental endstations,
    \item evaluated general-purpose lossless compression algorithms for reducing archival storage requirements, and
    \item characterized the scalability and throughput of the investigated compression methods.
\end{itemize}
Our study contributes to the development of efficient and scalable data management strategies for both PETRA III and the planned PETRA IV facility. PETRA IV was recently shortlisted as a priority research infrastructure by the German Federal Ministry of Research, Technology and Space. The insights gained from this work may also be relevant to the broader synchrotron and photon-science community and to initiatives such as LEAPS~\citep{LEAPS} and DAPHNE4NFDI~\citep{Dolcet02112023}, as well as to other large-scale light sources. These efforts ultimately support the continued advancement of scientific research at DESY and beyond. To our knowledge, a systematic cross-beamline evaluation of lossless compression methods for heterogeneous PETRA III datasets has not previously been reported.

We describe the background to compression methods in Sec.~\ref{sec:background}, the PETRA III data used in this study in Sec.~\ref{sec:corpus}, and our benchmarks and methodology in Sec.~\ref{sec:methods}. Finally, results are presented in Sec.~\ref{sec:results} and we discuss their implications in Sec.~\ref{sec:discussion}.

%%%%%%%%%%%%%%%%%%%%%%%%%%%%%%%%%%%%%%%%%%%%%%%%%%%%%%%%%%%%%%%%%%%%%%%%%%%%%%%%%%%%%%%%%%%%%%%%%%%%%%%%%%%%%%%%%%%%%%%%%%%%%%%%%%%%%%%%%%%%%%%%%%%%%%%%%%%%%%%%%%%%%%%%%%%%%%%%%%%%%%%%%%%%%%%%%%%%%%%%%%%%%%%
\section{Background and Related Work}
\label{sec:background}
Data compression methods are commonly categorized as either \emph{lossless} or \emph{lossy}. Lossless compression enables exact bit-wise reconstruction, whereas lossy compression permits some (potentially bounded) reconstruction error. Lossy approaches, including SZ \citep{Zhao2021}, zfp \citep{Lindstrom2014}, and others \citep{Gong2023}, often apply sophisticated mathematical models and have demonstrated excellent compression ratios. 
However, lossy compression methods have seen limited adoption within the synchrotron community due to the costly and difficult-to-reproduce nature of such experiments.
In contrast, general-purpose lossless algorithms such as Zstandard \citep{Collet2021}, Brotli \citep{Alakuijala2018}, and DEFLATE \citep{Deutsch1996} typically achieve lower compression ratios. These methods commonly rely on redundancy reduction and entropy encoding and can be applied to a broad range of numerical and non-numerical data types. To contextualize the variety of lossless compression tools, Tab.~\ref{tab:tools} summarizes the underlying algorithms and dominant compression principles represented in this study.

\begin{table*}
\centering
\caption{Overview of the compression tools evaluated in this study, Brotli~\citep{Alakuijala2018}, gzip~\citep{Deutsch1996gzip}, lrzip~\citep{Kolivas2008}, lzop~\citep{Oberhumer1996}, LZ4~\citep{Collet2011}, XZ Utils~\citep{LasseCollinXZ}, ZPAQ~\citep{Mahoney2011}, Zstandard~\citep{Collet2021}, and 7-Zip~\citep{Pavlov1999}, as well as their algorithms and backends, LZMA/LZMA2~\citep{Pavlov1999}, bzip2~\citep{Seward1998}, DEFLATE~\citep{Deutsch1996}, and PPMd~\citep{Shkarin2002}.}
\label{tab:tools}
\begin{tabularx}{\textwidth}{llX}
\toprule
Tool & Algorithm(s)/backend(s) & Main compression principle \\
\midrule
Brotli & Brotli & LZ77-style dictionary compression with statistical context modeling\\
gzip & DEFLATE & LZ77 dictionary matching and Huffman coding \\
lrzip & LZMA, LZO, bzip2, & Long-range redundancy reduction with backend compression \\
      & gzip, ZPAQ &\\
lzop & LZO & Fast LZ77-family dictionary compression \\
LZ4 & LZ4 & Very fast LZ77-family dictionary compression \\
XZ Utils & LZMA2 & Large-dictionary LZ-based compression and range coding \\
ZPAQ & ZPAQ & Context mixing and arithmetic coding for maximal compression.\\
Zstandard & Zstandard & LZ77-style dictionary matching with FSE/Huffman coding \\
7-Zip & LZMA, LZMA2, bzip2, & Dictionary and statistical context modeling \\
      & DEFLATE, PPMd &  \\
\bottomrule
\end{tabularx}
\end{table*}

Synchrotron experiments generate a wide variety of data types, including images and control data, stored in formats such as CBF \citep{Bernstein2006}, TIFF, and HDF5 \citep{hdf5}.
The NeXus standard \citep{Koennecke2015} has emerged as the most prominent method to store the multidimensional numeric arrays acquired during the experiments. 
Many of these formats natively support compression or can be extended through plugin mechanisms such as HDF5 filters. Common examples include LZ4 compression combined with Bitshuffle \citep{Masui2015}, as used by EIGER detectors \citep{Foerster2016}.
Beyond file-based storage, array database systems such as RasDaMan \citep{Reiner2002} and SciDB \citep{Brown2010} provide query capabilities for large-scale raster data. However, such systems remain less prevalent within the synchrotron user community than conventional file-based workflows.

While some authors have developed custom compression approaches for synchrotron data, these algorithms employ \emph{domain-specific} methods for certain data types \citep{Fu2021, Minxing2024} or measurement techniques \citep{Bernstein2025, Galchenkova2024}. To date, several researchers have presented benchmarks for lossless compression algorithms \citep{Chen2023, Gopinath2020, Gupta2017, Sun2025}, but none were specifically tailored to synchrotron sources. To the best of our knowledge, no comprehensive study of \emph{general-purpose} data compression efficacy has been presented for data from synchrotron facilities.

%%%%%%%%%%%%%%%%%%%%%%%%%%%%%%%%%%%%%%%%%%%%%%%%%%%%%%%%%%%%%%%%%%%%%%%%%%%%%%%%%%%%%%%%%%%%%%%%%%%%%%%%%%%%%%%%%%%%%%%%%%%%%%%%%%%%%%%%%%%%%%%%%%%%%%%%%%%%%%%%%%%%%%%%%%%%%%%%%%%%%%%%%%%%%%%%%%%%%%%%%%%%%%%
\section{Corpus Characterization\label{sec:corpus}}
We obtained read access to representative PETRA III experiments (\emph{beamtimes}) from beamlines p05, p07, p09, p10, and p11 through the responsible beamline scientists.
Together, these beamlines account for approximately 80.5\% of all data stored on the online filesystem prior to tape archival.
The resulting data corpus encompassed $>212$~TiB of data across 10 beamtimes from 2024 and 2025 (p05 -- one beamtime, p07 -- two beamtimes, p09 -- three beamtimes, p10 -- three beamtimes, p11 -- one beamtime). Each beamtime followed a directory structure imposed by the DESY computing center\footnote{\url{https://docs.desy.de/asap3/}}. We exclusively considered those folders that are automatically archived to tape -- \emph{raw} for raw data, \emph{processed} for processed data, and \emph{shared} for user-specific analysis files (e.g., scripts, intermediate data).

For characterization of the acquired dataset, we iterated over each of the considered directories per beamline (without following symlinks). Per unique suffix and directory, file counts and sizes were determined. Each file suffix was manually attributed to representative file categories.

The collected data required $212.28$ TiB of disk space across various experimental techniques, detector types, and built-in compression approaches from the participating beamlines, see Tab.~\ref{tab:exp_techniques}. Note that raw data was often already pre-compressed by built-in detector functionality (e.g., Bitshuffle+LZ4 HDF5 filters for Eiger, or CBF byte-offset compression for Pilatus detectors).

\begin{table*}
\caption{Experimental techniques, detector types, and built-in detector compression used to generate the raw files per considered beamline (p05, p07, ...) and beamtime (1-10).} \label{tab:exp_techniques}
\resizebox{\linewidth}{!}{
\begin{tabular}{@{}ll|lll@{}}
\toprule
\textbf{Beamline} & \textbf{Beamtime} & \textbf{Experimental Setup}                          & \textbf{Detector}                & \textbf{Built-in Compression} \\ \midrule
\textbf{p05}      & \textbf{1}        & Microtomography                                      & custom, based on CMOSIS CMV20000 & -                    \\
\textbf{p07}      & \textbf{2}        & X-ray Diffraction                                    & Pilatus 2M                       & Byte-offset CBF      \\
\textbf{}         & \textbf{3}        & X-ray Diffraction                                    & Eiger 4M                         & LZ4 + Bitshuffle     \\
\textbf{p09}      & \textbf{4}        & High-Throughput Pharmaceutical Screening with X-rays & Pilatus 6M                       & Byte-offset CBF      \\
\textbf{}         & \textbf{5}        & High-Throughput Pharmaceutical Screening with X-rays & Pilatus 6M                       & Byte-offset CBF      \\
\textbf{}         & \textbf{6}        & High-Throughput Pharmaceutical Screening with X-rays & Pilatus 6M                       & Byte-offset CBF      \\
\textbf{p10}      & \textbf{7}        & 4-circle SAXS/WAXS setup                             & Eiger 4M                         & LZ4 + Bitshuffle     \\
\textbf{}         & \textbf{8}        & 4-circle SAXS/WAXS setup                             & Eiger 4M, Eiger 500k             & LZ4 + Bitshuffle     \\
\textbf{} & \textbf{9} & Nanofocus setup (GINIX) & Eiger 4M, Pilatus 300k, pco-edge, zyla & LZ4 + Bitshuffle, Byte-offset CBF, - , - \\
\textbf{p11}      & \textbf{10}       & Serial Crystallography with Tapedrive                & Dectris Eiger 16M                & LZ4 + Bitshuffle     \\ \bottomrule
\end{tabular}
}
\end{table*}

Total storage volume per beamline was broadly comparable (min. $20.55$ TiB, max. $59.05$ TiB), but varied substantially between individual experiments (min. $249.49$ GiB, max. $57.04$ TiB), see Tab.~\ref{tab:corpus}.

\begin{table*}
\caption{Corpus composition by beamline (p05, p07, ...), beamtime (1-10), and file category (as manually assigned based on file suffix).\label{tab:corpus}}
\resizebox{\linewidth}{!}{
\begin{tabular}{@{}lllllll|l@{}}
\toprule
\textbf{Beamline} &
  \textbf{Beamtime} &
  \textbf{CBF} &
  \textbf{HDF5/NeXus} &
  \textbf{Image} &
  \textbf{Misc.} &
  \textbf{Text} &
  \textbf{Total} \\ \midrule
\textbf{p05}                  & \textbf{1}  & -          & 1.23 GiB   & 28.68 TiB  & 85.96 GiB  & 2.34 GiB   & \textbf{28.77 TiB}  \\
\multirow{2}{*}{\textbf{p07}} &
  \textbf{2} &
  821.60 GiB &
  2.31 GiB &
  50.82 MiB &
  117.69 GiB &
  52.43 MiB &
  \textbf{941.70 GiB} \\
                              & \textbf{3}  & -          & 19.63 TiB  & 5.12 GiB   & 431.37 MiB & 40.63 MiB  & \textbf{19.63 TiB}  \\
\multirow{3}{*}{\textbf{p09}} & \textbf{4}  & 435.62 GiB & 9.03 GiB   & -          & 995.74 MiB & 1.32 GiB   & \textbf{446.94 GiB} \\
                              & \textbf{5}  & 26.71 TiB  & 106.85 MiB & 2.82 GiB   & 60.86 GiB  & 30.83 TiB  & \textbf{57.60 TiB}  \\
                              & \textbf{6}  & 1.01 TiB   & -          & -          & 195.95 KiB & 12.70 MiB  & \textbf{1.01 TiB}   \\
\multirow{3}{*}{\textbf{p10}} & \textbf{7}  & -          & 52.88 GiB  & 196.41 GiB & 183.15 MiB & 13.15 MiB  & \textbf{249.49 GiB} \\
                              & \textbf{8}  & -          & 2.37 TiB   & 40.54 MiB  & 727.27 GiB & 57.82 MiB  & \textbf{3.08 TiB}   \\
                              & \textbf{9}  & 6.50 GiB   & 16.11 TiB  & 24.09 TiB  & 2.44 TiB   & 5.41 GiB   & \textbf{42.65 TiB}  \\
\textbf{p11}                  & \textbf{10} & -          & 57.83 TiB  & 34.04 MiB  & 8.11 GiB   & 104.72 GiB & \textbf{57.94 TiB}  \\ \midrule
\textbf{Total} &
  \textbf{} &
  \textbf{28.96 TiB} &
  \textbf{96.00 TiB} &
  \textbf{52.97 TiB} &
  \textbf{3.42 TiB} &
  \textbf{30.94 TiB} &
  \textbf{212.28 TiB} \\ \bottomrule
\end{tabular}
}
\end{table*}

\begin{figure*}
    \centering
    \subfloat[File size share per beamline and subfolder]{%
        \includegraphics[width=0.28\textwidth]{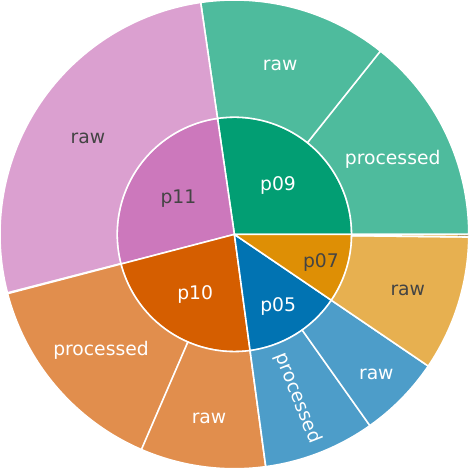}%
        \label{subfig:subdirectories}%
    }
    \hfill
    \subfloat[File counts per category]{%
        \includegraphics[width=0.71\textwidth]{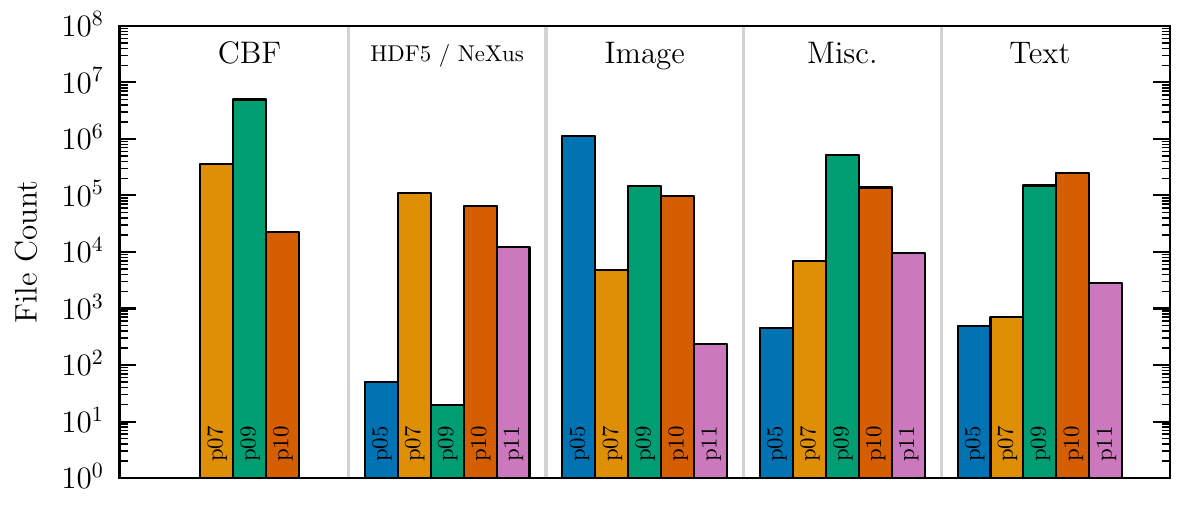}%
        \label{subfig:file_counts}%
    }

    \caption{Acquired file statistics across data corpus. \textbf{Left}: Share per subfolder and beamline (accumulated across all beamtimes). \textbf{Right}: Total file count per file category and beamtime (accumulated across all subfolders and beamtimes).}
    \label{fig:file_statistics}
\end{figure*}

Per beamline, each unique suffix was attributed to one of five categories (\emph{HDF5/NeXus}, \emph{Image}, \emph{Text}, \emph{CBF}, and \emph{Miscellaneous}). Note that the \emph{Image} category refers to individual image files (e.g., TIFF, PNG), in contrast to \emph{HDF5/NeXus}, where images are typically stored along an axis of a multi-dimensional (e.g., three-dimensional) array.
\emph{HDF5/NeXus} containers comprised the major part of the corpus ($96.00$ TiB, $45.2\%$ of total size), followed by \emph{Image} and \emph{Text} ($25.0\%$ and $14.6\%$, respectively). Considerable heterogeneity was observed between the major filetypes at individual beamtimes (e.g., p11 -- $99.81\%$ \emph{HDF5/NeXus}; p05 -- $99.7\%$ \emph{Image}).
Across all beamlines, we observed $64.6\%$, $35.4\%$, and $0.004\%$ of the data to reside in the directories for \emph{raw}, \emph{processed}, and \emph{shared} data, respectively. Interestingly, for beamlines p05, p09, and p10, \emph{processed} user data required approximately as much disk space as the \emph{raw} data (max. $62.4\%$ of beamtime on disk, cf. Fig.~\ref{subfig:subdirectories}).

Further analysis showed that file counts varied by several orders of magnitude between categories and beamlines, see Fig.~\ref{subfig:file_counts}. \emph{HDF5/NeXus} files were the least numerous ($\sim\!189,000$ files, 2.4\% of all files), whereas \emph{CBF} files dominated the corpus ($\sim\!5.3$ million files, 67.3\%). File sizes also varied substantially, with \emph{HDF5/NeXus} files being considerably larger on average than the other categories, see Tab.~\ref{tab:filecount}. Moreover, file sizes within each category spanned multiple orders of magnitude. While \emph{CBF} and \emph{Miscellaneous} files occurred primarily in the raw and processed directories, respectively, occurrence was approximately balanced for the other file types.

\begin{table*}
\centering
\caption{Percentage of file numbers, together with file size statistics, per directory and file category. \label{tab:filecount}}
\begin{tabular}{@{}llllll@{}}
\toprule
\textbf{Subdirectory} & \textbf{CBF} & \textbf{HDF5/NeXus} & \textbf{Image} & \textbf{Misc.} & \textbf{Text} \\ \midrule
\textbf{processed} & 0.06\%  & 28.46\% & 63.02\% & 94.73\% & 48.32\% \\
\textbf{raw}       & 99.94\% & 71.54\% & 36.91\% & 1.42\%  & 43.53\% \\
\textbf{shared}    & 0.00\%       & 0.00\%     & 0.07\%   & 3.85\%  & 7.15\%  \\ \midrule
\textbf{Min. Size} & 4.00 KiB  & 1.36 KiB & 0.00 Byte & 0.00 Byte & 0.00 Byte \\
\textbf{Avg. Size} & 5.68 MiB  & 532.99 MiB & 38.09 MiB & 2.01 MiB & 66.48 MiB \\
\textbf{Max. Size} & 6.98 MiB  & 61.59 GiB & 52.73 GiB & 244.20 GiB & 68.99 GiB \\ \bottomrule
\end{tabular}
\end{table*}
Of note, we observed strong variations between individual beamtimes (e.g., one of the three beamtimes from beamline p10 contains $80.8\%$ of all image files observed for that endstation).

%%%%%%%%%%%%%%%%%%%%%%%%%%%%%%%%%%%%%%%%%%%%%%%%%%%%%%%%%%%%%%%%%%%%%%%%%%%%%%%%%%%%%%%%%%%%%%%%%%%%%%%%%%%%%%%%%%%%%%%%%%%%%%%%%%%%%%%%%%%%%%%%%%%%%%%%%%%%%%%%%%%%%%%%%%%%%%%%%%%%%%%%%%%%%%%%%%%%%%%%%%%%%%%
\section{Methods}
\label{sec:methods}
Data compression is typically characterized by three performance metrics: compression ratio, compression throughput, and decompression throughput. No single compression algorithm simultaneously optimizes all three metrics, so one always has to make trade-offs. Depending on where compression is used in the data pipeline, the priorities may be different: For long-term archival, compression ratio and decompression throughput are typically prioritized to minimize storage requirements while maintaining efficient data access. Compression throughput is less critical in this context because archival data are generally compressed only once. In contrast, compression performed directly at the detector stage prioritizes throughput over compression ratio in order to avoid acquisition bottlenecks during data generation.

Motivated by these trade-offs, we formulate compressor selection as a multi-objective optimization problem. We aim to identify Pareto-efficient solutions, i.e., configurations for which no metric can be improved without degrading another. This will allow us to identify compression methods appropriate for different operational requirements. Note, however, that a full three-dimensional Pareto analysis is computationally expensive and unnecessary for the objectives of this study. First, the goal of this paper is not to exhaustively characterize the complete parameter space, but to be able to make informed and, most importantly, practical decisions. Second, the compression and decompression throughput are empirically observed to be correlated for most compressors. Furthermore, compression and decompression throughput exhibit opposite optimization directions, since bypassing compression entirely yields maximal effective throughput. It is therefore sufficient to break this problem down into two separate 2-dimensional Pareto optimizations: in the compression ratio vs compression throughput, and the compression ratio vs decompression throughput subspace. Because compression and decompression throughput are correlated for most algorithms evaluated in this study, the resulting two-dimensional Pareto fronts exhibit substantial overlap.

\subsection{Compression tools and algorithms}
We restricted our study to lossless, general-purpose compression tools with command-line interfaces to facilitate reproducible benchmarking and integration into automated archival workflows. Based on expected compression ratio, throughput, and prevalence within the user community, we selected nine representative tools, summarized in Tab.~\ref{tab:tools}.
Where applicable, multiple compression algorithms within a given tool were evaluated (e.g., LZMA2, bzip2, DEFLATE, and PPMd in 7-Zip).
The evaluated methods included both dictionary-based and statistical approaches, covering LZ77-type compressors (e.g., gzip, LZ4, Zstandard), large-dictionary methods (e.g., XZ, Brotli), and context-mixing models (e.g., ZPAQ). We also performed coarse-grained parameter scans over window sizes and a selection of other tunable parameters that each tool offers, following the procedure outlined in Sec.~\ref{sec:Optimization}. These scans were not intended to exhaustively optimize each compressor configuration, but rather to characterize practical trade-offs between compression ratio and throughput. A comprehensive summary of our compression configurations can be found in Appendix~\ref{sec:compressors}.

\subsection{Benchmarking Setup}
All compression tasks were performed on the \emph{maxwell} cluster at the DESY computing center. To ensure consistent benchmarking conditions, we selected dual-socket nodes with 2 x AMD EPYC 75F3 CPUs and 512 GB of RAM for our benchmarks. To compare results from different nodes, we furthermore restricted our tests to only run on nodes where the x2APIC extension was disabled in the BIOS, as we observed a performance variation of approximately 4-5\% depending on its state\footnote{Whether this extension is enabled or disabled does not make a difference ultimately as long as it is the same on all benchmarking nodes. We simply required x2APIC to be disabled, as this configuration was present on the majority of available nodes.}. We observed no further measurable difference between nodes. These nodes were selected because their large number of L3 cache domains and substantial memory capacity allowed us to isolate many single-threaded compression tasks and to keep both input and output data in memory, enabling rapid evaluation of the large configuration space considered in this study.

For benchmarking, we used approximately 2~GB of data from randomly selected files from one of the major file categories, \emph{CBF}, \emph{HDF5/NeXus}, \emph{Text}, or \emph{Image}. If necessary, multiple files with applicable suffixes were combined into \emph{tar}-archives of the desired size (max. 1,000 files). A payload size of 2~GB was chosen as a compromise between representativeness and practical benchmark runtime. We benchmarked each file category separately, as their compressibility can vary significantly. Compressibility depends both on the underlying data characteristics and on algorithm-specific optimizations, such as PPMd for textual data.

Each compression task was launched from a Python script using the subprocess module and timed using \texttt{/usr/bin/time}. For practical reasons, we timed out (de-)compression tasks that have a throughput of less than 1 MiB/s, which corresponds to about 30 minutes for a 2 GB file. Consequently, some configurations capable of achieving higher compression ratios may not be represented in our results. However, such methods would also be impractical for large-scale deployment due to their extremely low throughput.

\begin{figure}
    \centering
    \includegraphics[width=\linewidth]{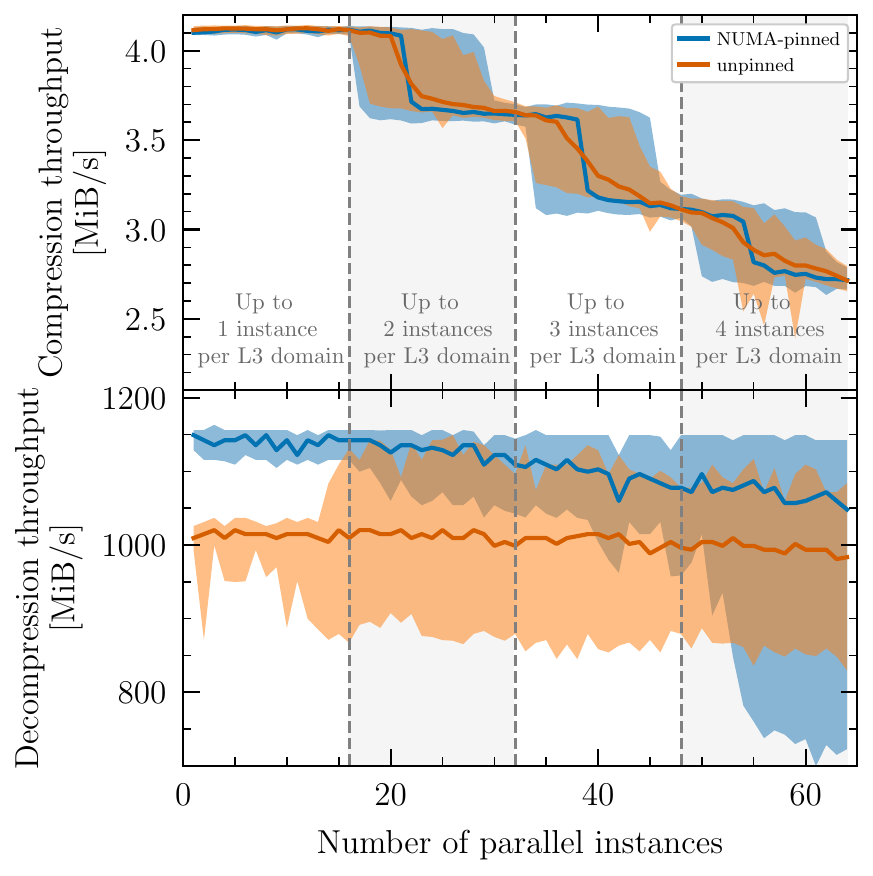}
    \caption{Median compression \textsl{(top)} and decompression \textsl{(bottom)} throughput for a varying number of single-threaded instances run in parallel on 2 x AMD EPYC 75F3 CPUs. The error bands represent the 5\% and 95\% percentiles across at least 64 repetitions. Labeled blocks indicate the number of instances sharing each L3 domain, which is enforced through pinning via \texttt{numactl}. For the unpinned configuration, L3 placement is determined by the OS scheduler rather than enforced; the similar step pattern nonetheless visible in the unpinned curve suggests that the scheduler distributes threads in a comparable way. The compression configuration uses Zstandard with compression level 16 and corresponds to our \textsl{Read-heavy archival} benchmark point summarized in Tab.~\ref{tab:summary}.}
    \label{fig:pinning}
\end{figure}

We confirmed explicitly that we can run up to 16 single-threaded compression tasks concurrently on these nodes without observing significant performance interference. This is achieved by pinning each compression task via \texttt{numactl} to the CPU cores associated with one of the 16 L3 cache domains available on each node. Pinning each task to a dedicated L3 cache domain proved sufficient to isolate workloads and reduce runtime variance. We furthermore created local NUMA-aware in-memory copies of the input files to minimize filesystem I/O effects during benchmarking, and directed all compression output to memory as well.

This is illustrated in Fig.~\ref{fig:pinning}, where we compare the median throughput of pinned and unpinned configurations for a varying number of parallel single-threaded compression instances. We used Zstandard configured for read-heavy archival work (see Tab.~\ref{tab:summary}), which lets us probe two extremes simultaneously: a slow, compute- and cache-bound compression stage and a bandwidth-limited, high-throughput decompression stage. The payload was a 2~GB tarball of \emph{CBF} files. We repeated the experiment until at least 64 individual measurements per data point were available, and confirmed that the results are qualitatively unchanged for other compressor configurations and payloads.

Because compression and decompression throughput differ by more than two orders of magnitude for this configuration, so do their respective hardware demands; this is reflected in Fig.~\ref{fig:pinning}. On the compression side, a clear step-like structure appears at multiples of 16, matching the number of available L3 domains: performance and timing spread degrade visibly each time an additional instance is placed in an already-occupied domain. The flat behavior up to 16 instances confirms that instances do not interfere as long as each occupies its own L3 cache domain. Because this configuration is compute- and cache-bound rather than limited by memory bandwidth or locality, NUMA placement is largely irrelevant here, and the pinned and unpinned versions perform almost identically.

On the decompression side, the difference between the pinned and unpinned versions is clearly visible, owing to the high throughput and correspondingly memory-access-intensive nature of decompression. Pinning each instance increases performance by roughly 15\% and reduces the timing variance, the latter being particularly important for our benchmarking procedure. Here too, performance remains stable as long as each instance has its own L3 domain, confirming that we can run up to 16 single-threaded instances in parallel without loss of throughput or increased runtime variance.

If not explicitly mentioned otherwise, we perform all benchmarking in single-threaded mode. Since not all compression methods support multi-threading, this enables a fair comparison across tools. In addition, single-threaded performance is particularly relevant for production-scale deployments, where it is often preferable to run multiple compression instances concurrently instead of relying on multi-threading. For block- or fragment-based compression methods, where block sizes are smaller than the total file size, the workload can typically be parallelized efficiently across independent compression tasks. In such settings, distributing work across multiple single-threaded processes can reduce synchronization and scheduling overhead compared to shared-memory multi-threading implementations. At sufficiently large scale, however, filesystem and memory I/O bandwidth can become the dominant bottleneck, limiting achievable throughput independently of the compression method itself.

\begin{figure}
    \centering
    \includegraphics[width=\linewidth]{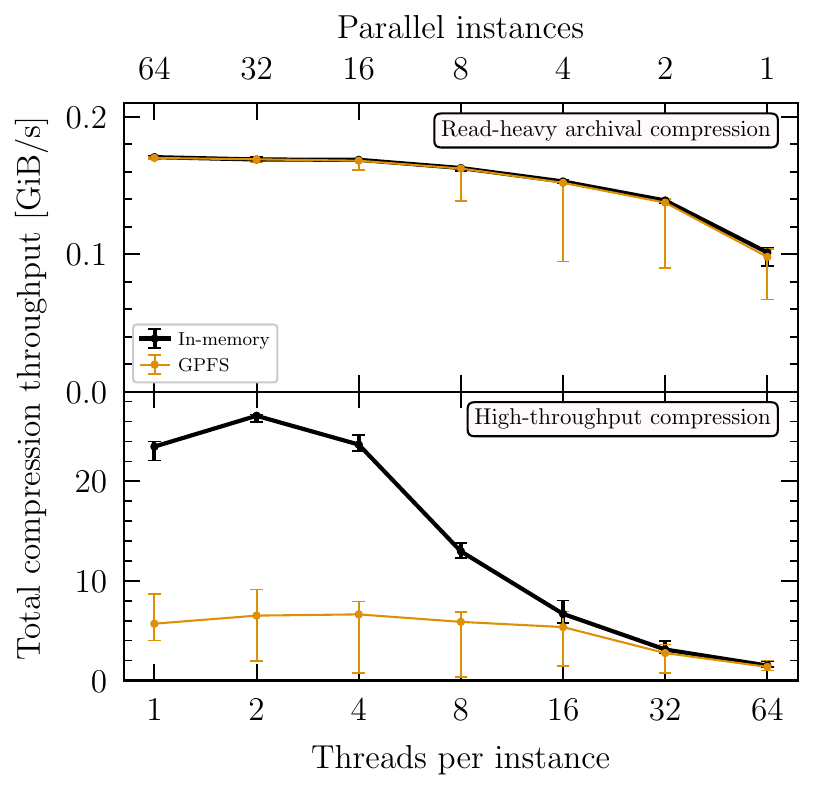}
    \caption{Compression throughput for a read-heavy archival \textsl{(top)} and a high-throughput \textsl{(bottom)} configuration based on Zstandard; see Tab.~\ref{tab:summary} for details. The payload was a 1~GB tarball of \emph{CBF} files, and the error bars represent the 5\% and 95\% percentiles. We compare I/O performance between an in-memory and a GPFS-backed setup, and vary the parallelization strategy from many independent single-threaded instances to a single 64-threaded instance, in each case fully utilizing the node.}
    \label{fig:threading}
\end{figure}

We illustrate these bandwidth effects in Fig.~\ref{fig:threading}, where we measure the total compression throughput per node under full utilization. We vary the parallelization approach from 64 instances with 1 thread each, through 32 instances with 2 threads, to a single instance with 64 threads. We repeated this experiment for the read-heavy archival and the high-throughput compressor configurations given in Tab.~\ref{tab:summary}, applied to a 1~GB tarball from the \emph{CBF} file category, until at least 8 measurements per data point were available.

We deliberately omit this experiment for the other two configurations in Tab.~\ref{tab:summary}, as neither parallelizes well. For the archival ceiling configuration, ZPAQ distributes work by splitting the data into fragments and assigning one thread per fragment. Since our configuration is optimized for maximum compression ratio, however, we use fragments that are as large as possible to enlarge the context window, so that in practice only one or two fragments exist in total and ZPAQ cannot make effective use of multiple threads. The compression-bound staging configuration, on the other hand, relies on 7-Zip, and while 7-Zip supports multi-threading in general, its PPMd backend runs serially. Thus, this configuration likewise does not benefit from multi-threading.

The two remaining configurations are based on Zstandard, which does not suffer from these limitations for compression. For decompression, however, Zstandard decodes each stream on a single core, so throughput there is governed by the number of concurrent instances rather than by threads, and we therefore focus on compression only in the following. Even for Zstandard compression, the results in Fig.~\ref{fig:threading} reveal several subtleties that complicate the effective use of multi-threading in practice.

For the read-heavy archival configuration, compression is compute- and cache-bound, and largely insensitive to the thread--instance split, aside from a moderate drop at the highest thread counts. The in-memory and GPFS results remain nearly indistinguishable, indicating that the filesystem is not a bottleneck in this slow, compute- and cache-bound regime.

For the high-throughput configuration, compression throughput degrades sharply as work is consolidated into fewer, more heavily threaded instances, peaking at two threads per instance before falling off. Unlike the compute- and cache-bound archival case, throughput here is limited by memory and storage bandwidth, which is used more efficiently by many independent instances than by threading within a single one. This is also reflected in the pronounced GPFS gap for this configuration: the GPFS curve plateaus well below the in-memory values and converges with them only under heavy consolidation.

Overall, we conclude that the effective use of multi-threading for compression involves a number of non-obvious trade-offs and must be approached with care. In many cases, simply running multiple independent instances is preferable, both because some configurations cannot exploit threads at all and because bandwidth-bound workloads scale better across instances than within them.

\subsection{Optimization procedure}
\label{sec:Optimization}
Due to the large number of possible compressor configurations, we do not evaluate all of them exhaustively. Instead, we employ a Pareto-guided local neighborhood search. For each compression method, we begin with its default presets for multiple compression levels, which serve as initial candidates. We evaluate these configurations in terms of compression throughput, decompression throughput, and compression ratio, to construct two preliminary Pareto fronts.

We then identify configurations that are near Pareto-optimal on either front. Here, \textsl{near Pareto-optimal} refers to configurations lying within 5\% of the current Pareto front. For each such configuration, we explore local variations across all tunable parameters, including multi-parameter combinations but avoiding configurations we have already tested.

This process is repeated iteratively: whenever a new configuration shifts the Pareto front, its neighborhood is explored in the same manner. The search terminates once the Pareto front stabilizes and no additional near-Pareto-optimal configurations are identified.

We repeat this procedure separately for each compression mode within a given tool to capture distinct compression behaviors. As this optimization procedure requires evaluating more than $10^5$ individual compression settings in total, we use a single 2 GB benchmark file from each file category for this stage of the analysis. Since a single benchmark file may not perfectly represent the statistical properties of the full dataset category, small shifts in absolute compression ratio relative to the full dataset are expected. However, because all compressor settings are evaluated on the same benchmark file, the impact on relative comparisons between compressor configurations is expected to be limited, particularly for identifying Pareto-optimal operating points.

To improve representativeness, we constructed each benchmark file from the beamtime contributing the largest total storage volume within the respective category: \emph{raw} data from beamtime 10 for the \emph{HDF5/NeXus} category, \emph{raw} data from beamtime 1 for the \emph{Image} category, and \emph{raw} and \emph{processed} data from beamtime 10 for the \emph{CBF} and \emph{Text} categories, respectively.

We subsequently reevaluated a subset of the Pareto-dominant configurations on substantially larger datasets to improve statistical robustness. We did not repeat this procedure for all Pareto-dominant configurations, as many neighbouring operating points correspond to very similar compression settings and exhibit nearly identical performance characteristics.

Instead, for each file category, we selected twelve representative configurations from the compression Pareto front, twelve from the decompression Pareto front, twelve from the subset of the compression Pareto front located near the decompression-optimal region, and twelve from the corresponding decompression configurations near the compression-optimal region. After removing overlapping configurations between these selections, 32 configurations remained for the \emph{CBF} category, 40 for the \emph{HDF5/NeXus} category, 33 for the \emph{Image} category, and 36 for the \emph{Text} category.

For each selected configuration, we compressed up to sixteen 2 GB chunks, each assembled from randomly chosen files, from each major subdirectory (\emph{raw} / \emph{processed}) and beamtime whenever at least 20 GB of data were available. The final evaluation, therefore, used a total of 160 GB for the \emph{CBF} category, 208 GB for \emph{HDF5/NeXus}, 102 GB for \emph{Images}, and 74 GB for \emph{Text} files per configuration. Statistical aggregation was weighted according to the total storage volume represented by each directory.

%%%%%%%%%%%%%%%%%%%%%%%%%%%%%%%%%%%%%%%%%%%%%%%%%%%%%%%%%%%%%%%%%%%%%%%%%%%%%%%%%%%%%%%%%%%%%%%%%%%%%%%%%%%%%%%%%%%%%%%%%%%%%%%%%%%%%%%%%%%%%%%%%%%%%%%%%%%%%%%%%%%%%%%%%%%%%%%%%%%%%%%%%%%%%%%%%%%%%%%%%%%%%%%
\section{Results}
\label{sec:results}
\subsection{Comparison between tools}
\begin{figure*}
    \centering
    \includegraphics[width=\textwidth]{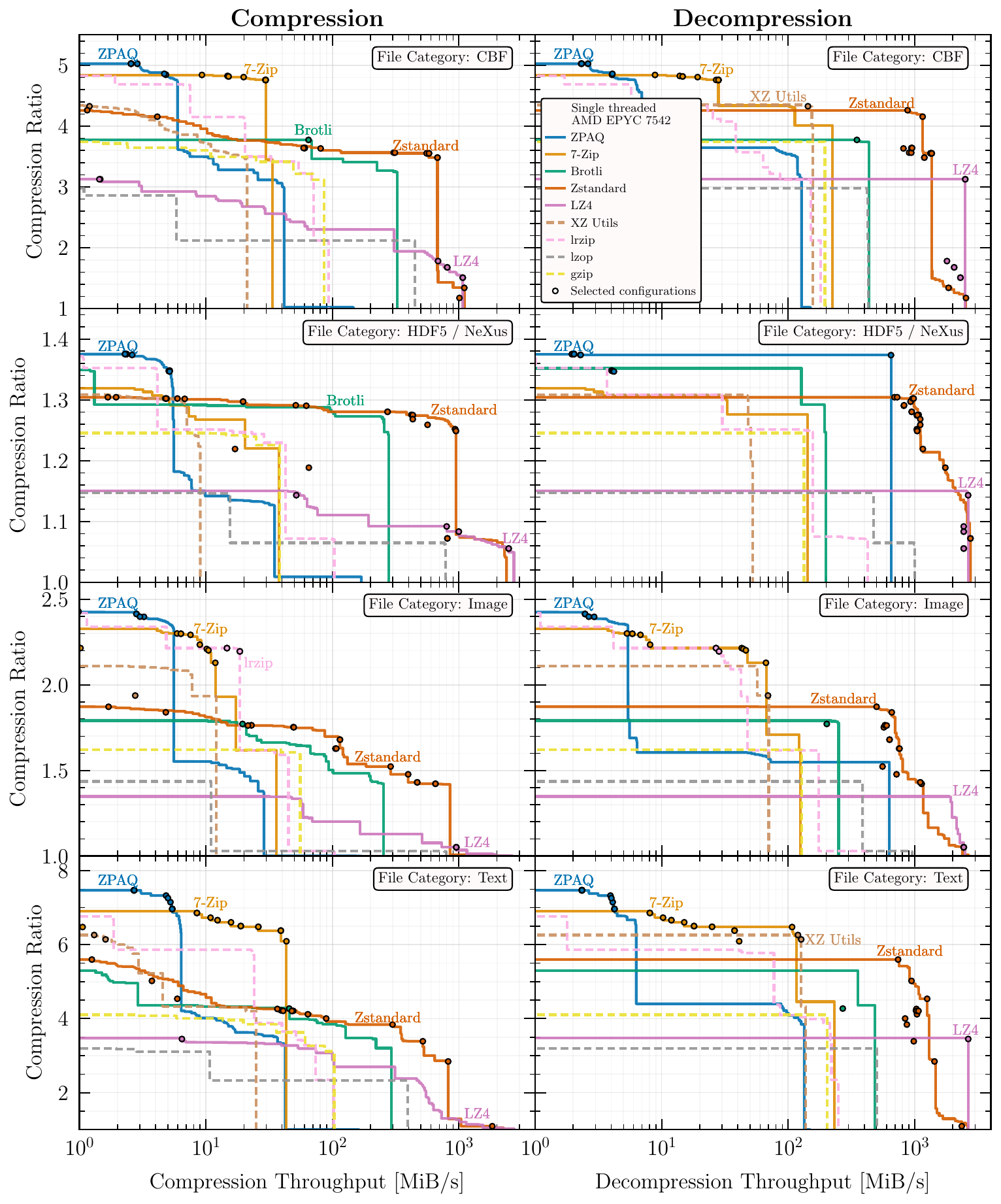}
    \caption{Compression \textsl{(left column)} and decompression \textsl{(right column)} Pareto fronts for all four major file categories \textsl{(top to bottom: \emph{CBF}, \emph{HDF5/NeXus}, \emph{Image}, and \emph{Text})}. We show all nine tools individually to highlight their respective trade-offs, while combining all tested algorithms and operating modes within a given tool into a single sample. Each compression configuration was benchmarked on a 2~GB file, and throughput was measured single-threaded on an AMD EPYC 75F3 CPU. Highlighted markers indicate the configurations selected for further analysis.}
    \label{fig:pareto_tools}
\end{figure*}

We applied the optimization procedure described in Sec.~\ref{sec:Optimization} to construct compression and decompression Pareto fronts separately for each compression tool and major file category. The resulting fronts are shown in Fig.~\ref{fig:pareto_tools}. Although the optimization was performed independently for each compression algorithm or operating mode within a tool, for example, bzip2, LZMA2, and PPMd within 7-Zip, all evaluated configurations belonging to the same tool were combined to obtain a unified Pareto front representation.

Across all file categories, the Pareto fronts separate naturally into distinct compressor families characterized by different algorithmic design priorities. In the high-throughput regime, Zstandard and LZ4 consistently dominate, frequently achieving compression and decompression throughputs exceeding 1~GiB/s. This behavior is expected, as both compressors were explicitly designed for low-latency operation using lightweight LZ77-style dictionary matching, low-overhead entropy coding, and highly optimized decoder implementations. While LZ4 yields the highest throughputs overall, the corresponding compression ratios are often relatively modest. Zstandard generally occupies a more balanced operating region, sacrificing a limited amount of throughput in exchange for substantially improved compression ratios. In practice, this may make Zstandard the more attractive choice when both throughput and storage efficiency are relevant considerations.

At the opposite end of the Pareto front, where compression ratio is prioritized over throughput, ZPAQ consistently achieves the strongest compression performance across all file categories. This observation is likewise unsurprising given the design philosophy of ZPAQ, which prioritizes maximal compression ratio through computationally expensive statistical modeling techniques, including context mixing and arithmetic coding. These approaches exploit long-range statistical structure beyond the capabilities of conventional dictionary-based compressors, albeit at the cost of substantially reduced throughput.

Between these two extremes, the Pareto structure becomes more dependent on the underlying data characteristics. Several compressors, most prominently 7-Zip, Brotli, lrzip, and XZ Utils, intermittently occupy Pareto-optimal regions whenever a compromise between throughput and compression ratio is sought. In this intermediate regime, the interaction between compressor architecture and data structure becomes particularly important. For example, 7-Zip dominates substantial parts of the Pareto front for the \emph{Text} category due to its implementation of the PPMd algorithm. Unlike dictionary-based compressors, PPMd employs probabilistic context modeling to predict symbol sequences, which is particularly effective for structured textual data exhibiting strong local symbol correlations. In contrast, compressors such as lzop and gzip do not appear competitive within the explored throughput--compression trade-off space for the datasets considered in this study.

\begin{figure*}[tbp]
    \centering
    \includegraphics[width=\textwidth]{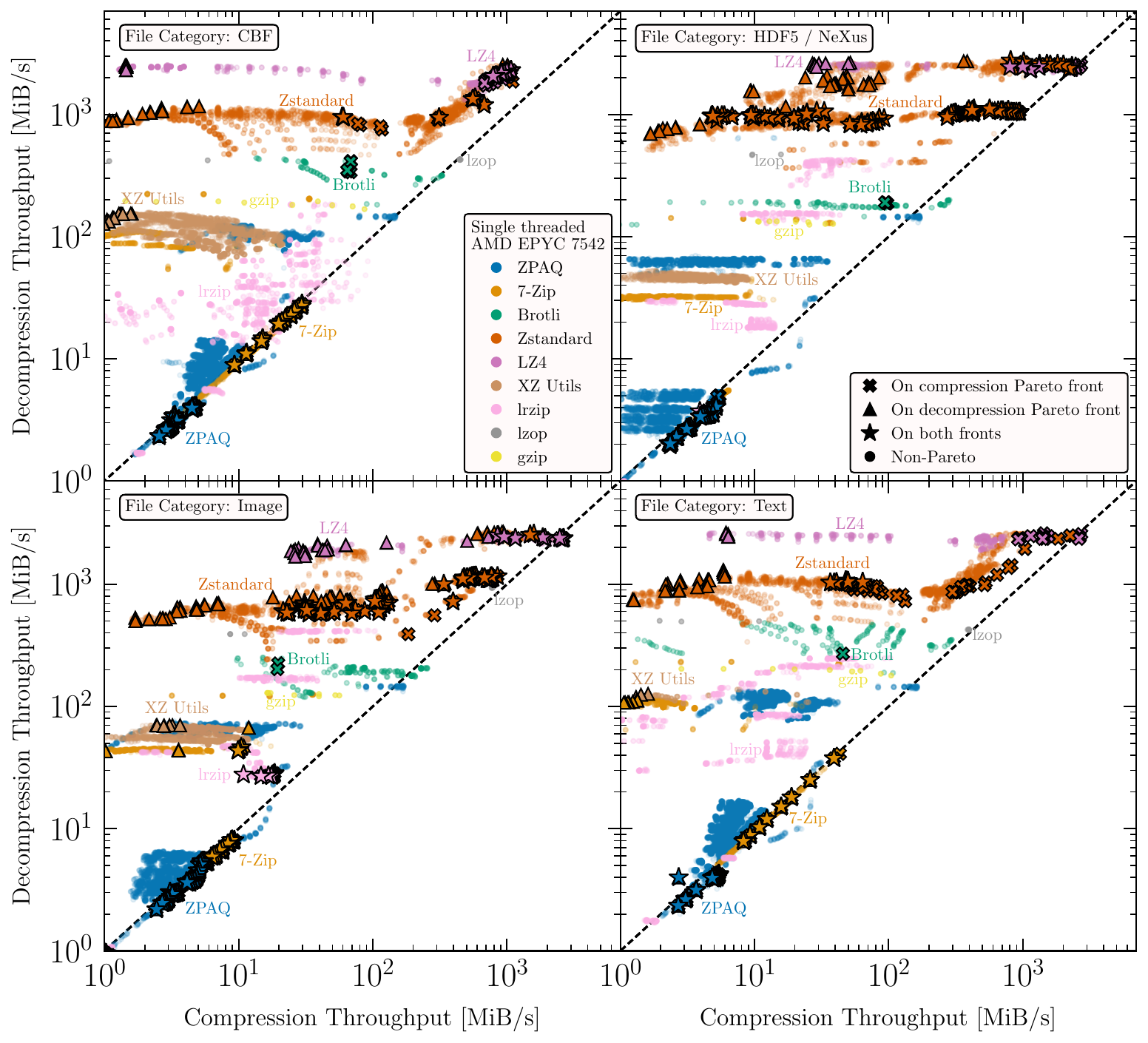}
    \caption{Compression and decompression throughput for all four major file categories \textsl{(clockwise: \emph{CBF}, \emph{HDF5/NeXus}, \emph{Text}, and \emph{Image})}. Shown are point clouds for each of the nine compressor tools, where the dataset comprises every compressor configuration evaluated during the optimization procedure. Each configuration was benchmarked on a 2~GB file, and throughput was measured single-threaded on an AMD EPYC 75F3 CPU. Highlighted markers indicate configurations located on the global compression Pareto front (\ding{54}), decompression Pareto front ($\blacktriangle$), or simultaneously on both fronts ($\bigstar$). The dashed line indicates equal compression and decompression throughput.}
    \label{fig:throughput}
\end{figure*}

We furthermore analysed the relation between compression and decompression throughput, shown in Fig.~\ref{fig:throughput}. Across most compressor families, we observe a substantial correlation between both quantities. In general, decompression throughput exceeds compression throughput, often by a considerable margin, reflecting the fact that many modern compression algorithms intentionally concentrate computational complexity in the encoder while maintaining comparatively lightweight decoder paths. This design principle is particularly evident for high-throughput compressors such as LZ4 and Zstandard, but is also visible for more compression-oriented methods. Only in the regime where compression and decompression throughput become comparable does decompression throughput begin to scale approximately proportionally with compression throughput. Across all evaluated configurations, compression throughput rarely exceeds decompression throughput.

\subsection{Comparison between datasets}
\begin{figure*}
    \centering
    \includegraphics[width=0.99\textwidth]{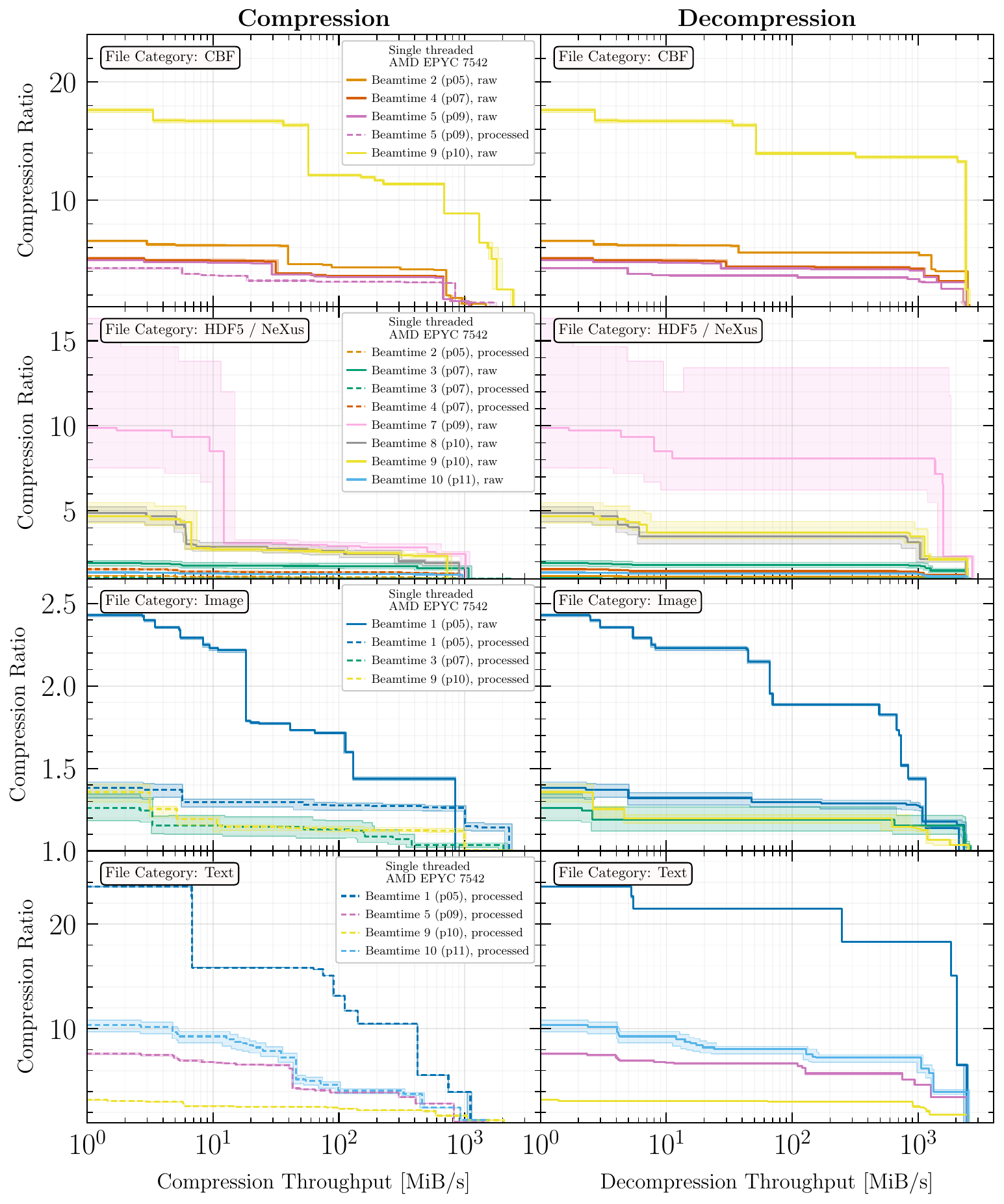}
    \caption{Compression \textsl{(left)} and decompression \textsl{(right)} Pareto fronts for all four major file categories using a selected subset of compressor configurations. Results are shown separately for each beamtime and subdirectory (\emph{raw} / \emph{processed}), illustrating the heterogeneity of the datasets. Each configuration was benchmarked on a representative set of sixteen 2~GB files. Throughput measurements were obtained single-threaded on an AMD EPYC 75F3 CPU. Shaded bands indicate the 25\% and 75\% quantiles in compression ratio and throughput around the mean values. Two datasets exhibiting unusually large compression ratios are omitted here for readability and shown separately in Fig.~\ref{fig:pareto_special_datasets}.}
    \label{fig:pareto_datasets}
\end{figure*}

\begin{figure*}
    \centering
    \includegraphics[width=\textwidth]{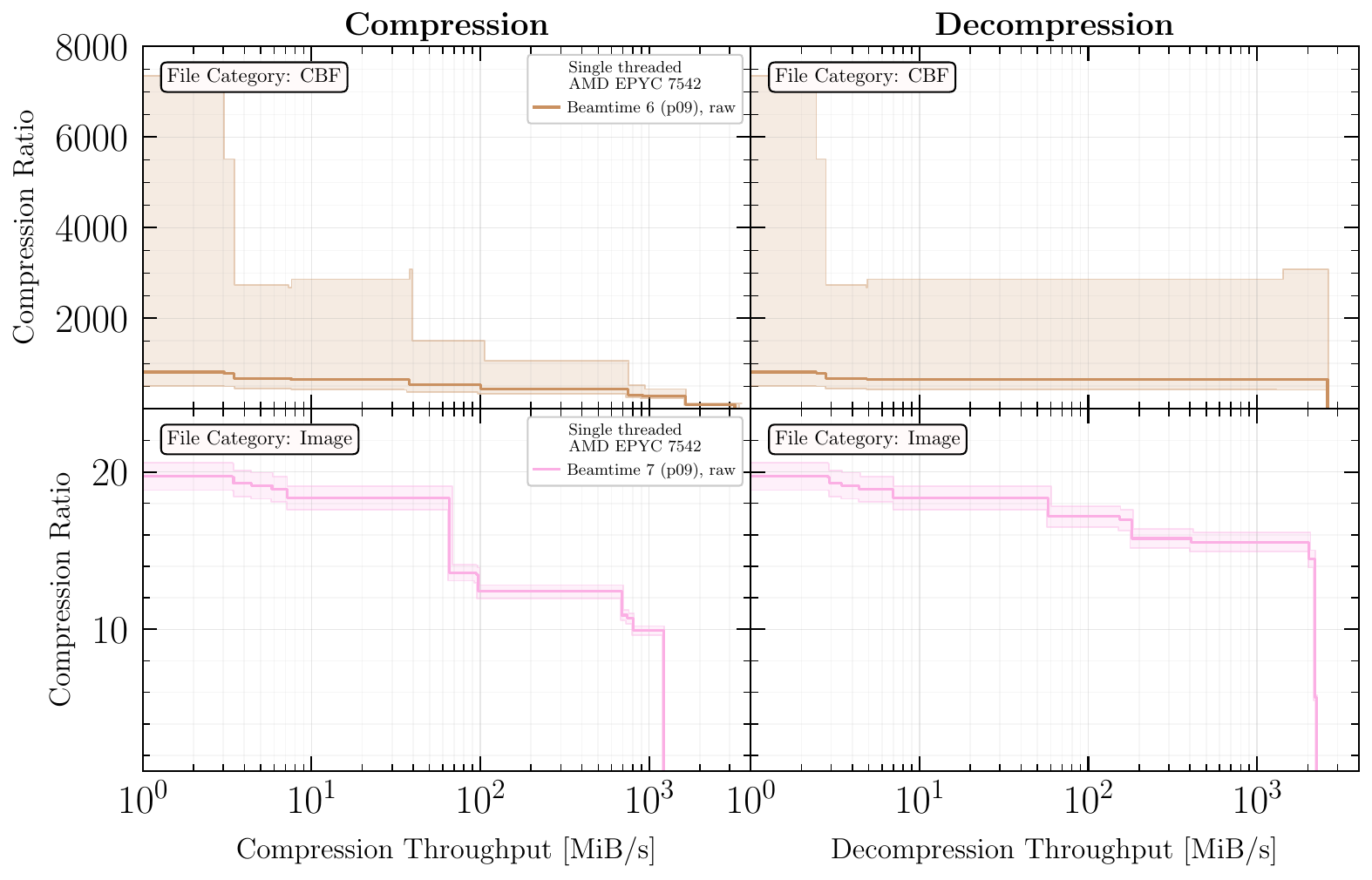}
    \caption{As Fig.~\ref{fig:pareto_datasets}, but for two datasets exhibiting unusually large compression ratios in the \emph{CBF} and \emph{Image} categories, respectively.}
    \label{fig:pareto_special_datasets}
\end{figure*}

In the previous subsection, all compressor configurations were benchmarked using a single representative 2~GB file for each major file category. As discussed in Sec.~\ref{sec:Optimization}, this initial optimization stage serves primarily to identify Pareto-optimal compressor configurations efficiently. To evaluate how robust these observations are across different experiments and data acquisition modes, we subsequently reevaluated a subset of the Pareto-optimal configurations on a substantially broader collection of datasets. The selected configurations are highlighted explicitly in Fig.~\ref{fig:pareto_tools}. Results obtained from this second benchmarking stage are shown in Fig.~\ref{fig:pareto_datasets} and Fig.~\ref{fig:pareto_special_datasets}.

The resulting Pareto fronts reveal substantial heterogeneity both between and within datasets. Most prominently, the achievable compression ratio varies by multiple orders of magnitude across different beamtimes, even within the same file category. In the \emph{CBF} category, for example, the maximal compression ratio ranges from approximately 4.3 for \emph{processed} data from beamtime~5 to more than 800 for \emph{raw} data from beamtime~6. Similarly large variations are observed for other categories, such as \emph{Image} files, where maximal compression ratios span from approximately 1.3 for beamtime~3 to around 20 for beamtime~7.

These differences primarily reflect variations in the statistical structure and redundancy of the underlying experimental data. In some cases, datasets contain substantial repeated or low-entropy regions that can be exploited effectively even by comparatively simple compression methods. In other cases, little residual redundancy remains available to general-purpose compressors. Existing detector-side or file-format-level compression can further reduce the achievable gains during archival compression, particularly for image-oriented formats that already incorporate lossless compression internally. The same holds for the \emph{HDF5/NeXus} category, where some but not all beamtimes already use gzip compression internally, explaining the poor compressibility observed in Fig.~\ref{fig:pareto_datasets}.

Note, however, that detector-side compression does not necessarily remove all redundancy present in the data. An illustrative example is the raw CBF dataset from beamtime 6 shown in Fig.~\ref{fig:pareto_special_datasets}, which achieves an average compression ratio exceeding 800 with ZPAQ despite already being compressed at the detector level. This detector employs byte-offset compression, a predictive encoding scheme that stores differences between neighbouring pixel values rather than the values themselves. While this approach is highly effective for reducing the dynamic range of crystallographic images and typically achieves compression ratios approaching two, it does not perform entropy coding. Consequently, highly repetitive structures can remain in the encoded data. In the beamtime 6 dataset, many neighbouring pixels are identical, resulting in long sequences of zero-valued offsets. These residual redundancies can subsequently be exploited very effectively by statistical compressors such as ZPAQ, leading to the exceptionally large compression ratios observed.

Beyond differences between experiments, several datasets also exhibit pronounced inhomogeneity within the same beamtime. This behavior is visible in the width of the 25\% and 75\% quantile bands on the compression ratio shown in Fig.~\ref{fig:pareto_datasets} and Fig.~\ref{fig:pareto_special_datasets}. For some datasets, these bands remain narrow across the entire Pareto front, indicating relatively homogeneous compressibility across files. In contrast, other datasets exhibit extremely broad quantile ranges, demonstrating that compressibility can vary substantially even within a single experiment. This observation suggests that dataset composition can influence achievable compression performance almost as strongly as the choice of compression algorithm itself.

While achievable compression ratios vary substantially between datasets, throughput measurements remain comparatively stable across beamtimes and file collections. This suggests that compression throughput is primarily governed by algorithmic complexity, whereas compression ratio depends much more strongly on the dataset-specific statistical structure.

Where direct comparisons between \emph{raw} and \emph{processed} datasets are possible, \emph{raw} data frequently exhibits higher compressibility. This may reflect the fact that scientific processing pipelines partially remove or decorrelate structural redundancies present in detector-level data.

\subsection{Extrapolated results}
The beamtime-specific Pareto fronts furthermore allow us to estimate compression performance for both individual file categories and the complete dataset. To obtain aggregate compression ratios for a single file category, we combine results from different beamtimes and subdirectories using weights proportional to their total storage contribution. Importantly, compression ratios are not averaged directly, as this would bias the result toward smaller benchmark subsets. Instead, we aggregate compressed and uncompressed file sizes separately and compute the effective compression ratio from their weighted sums, $\sum(w_i s_i)/\sum(w_i c_i)$. Here,  $s_i$ and $c_i$ denote the total uncompressed and compressed sizes of the benchmarked files associated with beamtime and subdirectory $i$. The weights are defined as $w_i= S_i / s_i$, where $S_i$ is the total storage volume represented by the corresponding beamtime and subdirectory $i$ in the full dataset. The combined throughput is computed analogously. This is done for each compressor configuration independently.

For the full-dataset analysis, we additionally consider heterogeneous compression strategies in which different file categories are compressed using different compressor configurations. This reflects a more realistic operational scenario, as the optimal compression method can depend strongly on the underlying data type, given the heterogeneity we observed between file categories. Allowing different file categories to use different compressor configurations can substantially improve the achievable global Pareto front compared to any single uniform compression strategy. To evaluate such mixed strategies, we compute the aggregate compression performance for all possible combinations of selected configurations across file categories and subsequently determine the corresponding global Pareto front.

\begin{figure*}
    \centering
    \includegraphics[width=\textwidth]{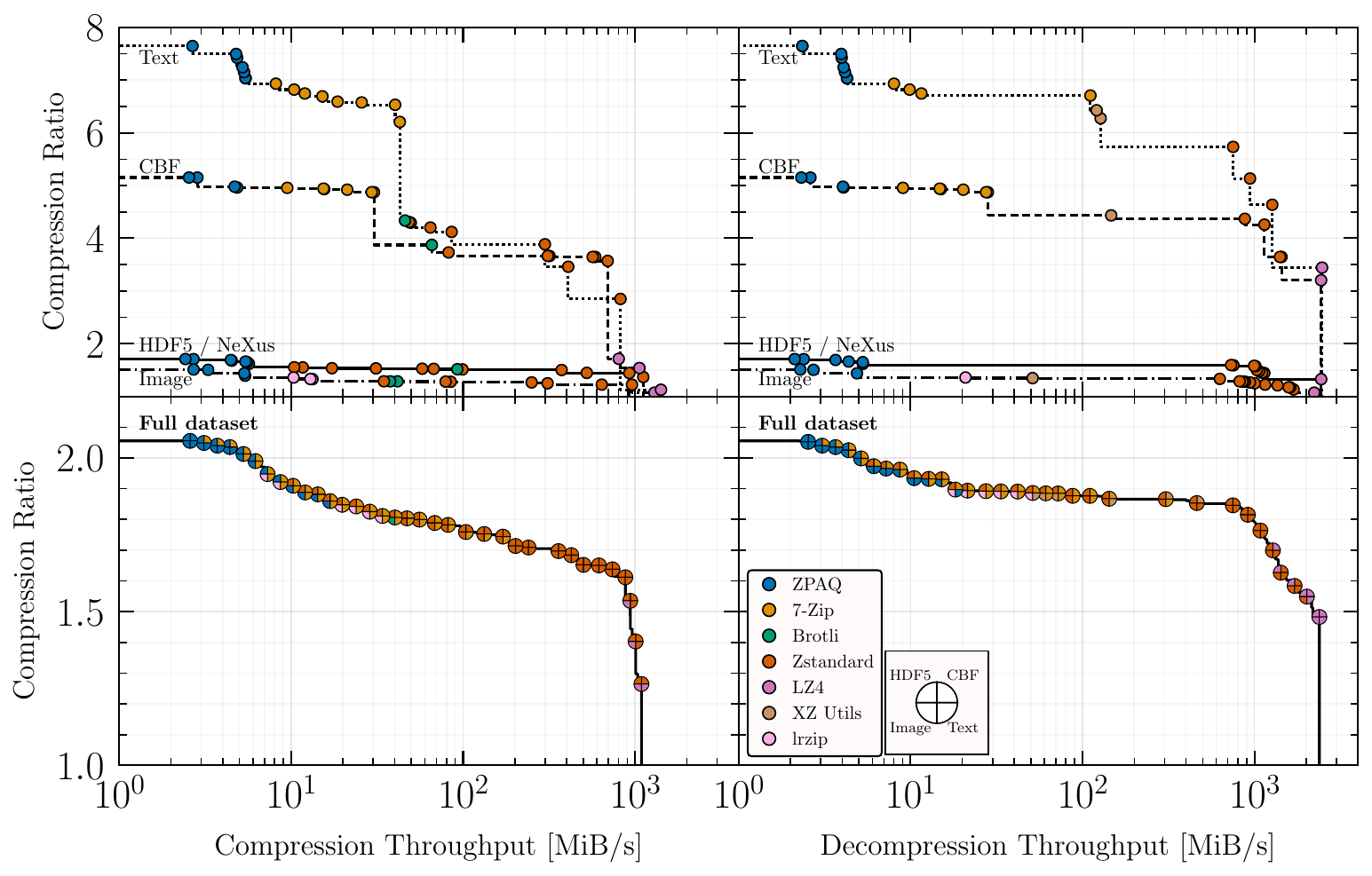}
    \caption{Compression \textsl{(left)} and decompression \textsl{(right)} Pareto fronts for all four major file categories using a selected set of compressor configurations. In addition, we present a combined extrapolated result if the entire dataset were to be compressed. Each compression configuration was benchmarked on a representative set of 2 GB files from all relevant beamtimes. Throughput was measured on an AMD EPYC 75F3 CPU using a single thread. Colored markers along the Pareto front indicate which compressor configuration(s) are used.
    }
    \label{fig:pareto_global}
\end{figure*}

We present the aggregate result in Fig.~\ref{fig:pareto_global}. It is apparent that \emph{CBF} and \emph{Text} categories overall benefit the most from data compression, with a maximum compression ratio of 5.2 and 7.6, respectively. \emph{HDF5/NeXus} and \emph{Image} files, on the other hand, only achieve a maximum compression ratio of around 1.7 and 1.5, which subsequently drags down the maximum compression ratio for the entire dataset to 2.1. The comparatively modest gains for \emph{HDF5/NeXus} and \emph{Image} categories indicate that these formats already employ effective built-in compression or otherwise contain limited residual redundancy available to general-purpose compressors. On the throughput side, the result is more homogeneous: Compression throughputs can be as high as 1 GiB/s, though more realistic configurations would probably be around a few hundred MiB/s. Decompression throughputs exhibit a comparatively sharp upper bound around 2--3~GiB/s.

For the complete dataset, the global Pareto front illustrates the achievable trade-offs between storage reduction and throughput.
Overall, we find for the entire dataset that a compression ratio of 2.1 at $\sim2$ MiB/s throughput, or 1.6 at $\sim900$ MiB/s, is achievable.

%%%%%%%%%%%%%%%%%%%%%%%%%%%%%%%%%%%%%%%%%%%%%%%%%%%%%%%%%%%%%%%%%%%%%%%%%%%%%%%%%%%%%%%%%%%%%%%%%%%%%%%%%%%%%%%%%%%%%%%%%%%%%%%%%%%%%%%%%%%%%%%%%%%%%%%%%%%%%%%%%%%%%%%%%%%%%%%%%%%%%%%%%%%%%%%%%%%%%%%%%%%%%%%
\section{Discussion and Outlook}
\label{sec:discussion}

\begin{table*}%[ht!]
\centering
\caption{Promising compressor configurations for a few different use cases based on our extrapolated result for the full dataset. All throughputs were measured on an AMD EPYC 75F3 CPU using a single thread.}
\label{tab:summary}
\begin{tabularx}{\textwidth}{c  c  c  X }
\toprule Compression & Compression & Decompression & Compressor Configuration \\
      Ratio   & Throughput & Throughput &  \\\midrule
\multicolumn{4}{l}{\textbf{Archival ceiling} (maximum compression ratio, rarely practical):} \\
2.1 & 2.5 MiB/s & 2.2 MiB/s & ZPAQ with BWT, ci1mst context model, and large block and fragment size.\\\\

\multicolumn{4}{l}{\textbf{Read-heavy archival} (compress once, decompress often, decent ratio):} \\
1.8 & 6.1 MiB/s & 971 MiB/s & Zstandard with compression level around 16.\\\\

\multicolumn{4}{l}{\textbf{Compression-bound staging} (ratio matters, but compression must not bottleneck data acquisition):} \\
1.8 & 55 MiB/s & 105 MiB/s & \emph{HDF5/NeXus} and \emph{Image}: Zstandard with compression level around 10. \emph{Text} and \emph{CBF}: 7-Zip with PPMd and compression level around 7.\\\\

\multicolumn{4}{l}{\textbf{High-throughput} (e.g. for interactive workflows):} \\
1.6 & 878 MiB/s & 1180 MiB/s & Zstandard with \texttt{--fast} and compression levels of $150$ or higher. \\\\

\end{tabularx}
\end{table*}

The increasing data volume generated by large-scale research facilities mandates the implementation of efficient data management strategies at the associated data centers. 
Driven by this motivation, we present the first comprehensive study of data heterogeneity and lossless general-purpose compression efficacy at the synchrotron radiation source PETRA III.

The study collected a corpus of representative data from major experimental endstations at PETRA III, encompassing ten experiments with more than 212 TiB of data. 
The survey confirmed significant heterogeneity among beamlines as well as among individual beamtimes, with \emph{HDF5/NeXus} and \emph{Image} files being predominant. 
In practice, such heterogeneity prevents data centers from fully leveraging synergies between beamlines.
Moreover, we observed that \emph{processed} user data required as much disk space as the \emph{raw} data in some cases, suggesting that data reduction needs to consider both \emph{processed} and \emph{raw} data. Close interaction with both user communities and infrastructure operations experts will therefore be critical for developing successful, holistic data reduction procedures.

Extrapolating from the benchmarked datasets, we found that high compression ratios are primarily enabled by the \emph{CBF} and \emph{Text} categories, whereas \emph{HDF5/NeXus} and \emph{Image} files exhibit substantially lower residual compressibility. In Tab.~\ref{tab:summary} we summarize several practical compressor configurations for archival and live-workflow scenarios that balance throughput against compression ratio.

The strong heterogeneity observed between beamlines and even individual beamtimes suggests that future studies should include larger and more diverse datasets. Since PETRA III and its associated beamlines continuously evolve, the present work necessarily represents a snapshot in time, and compression characteristics may change in the future. Nevertheless, the presented analysis provides a quantitative basis for evaluating archival compression strategies at PETRA III.

For throughput-oriented deployments, Zstandard and LZ4 consistently occupy the high-throughput region of the Pareto front, whereas ZPAQ dominates the high-compression regime. More generally, the results indicate that heterogeneous compression strategies, in which different file categories are compressed using different methods, can provide superior trade-offs compared to a single uniform compression policy.

Any production deployment should carefully consider integration with existing archival and retrieval workflows. Given the approximately 380 MiB/s uncompressed write bandwidth of an LTO-9 tape drive, many high-compression configurations would require parallel execution across multiple files to avoid becoming the archival bottleneck. We found that running multiple independent compression instances, rather than relying on multi-threading within a single instance, is often the better strategy, and that pinning each instance to a dedicated cache domain further improves throughput stability. Alternatively, hardware-accelerated approaches, including FPGA-based implementations~\cite{Chen2021, Bartik2015}, may offer a path toward combining higher throughput with stronger compression ratios.

In summary, the presented study demonstrates that compression performance at PETRA III is governed as much by dataset composition as by compressor choice. The resulting trade-offs between compression ratio and throughput provide a quantitative foundation for future archival and storage decisions and highlight the value of workload-aware compression strategies for large-scale scientific facilities.

\backmatter
%%%%%%%%%%%%%%%%%%%%%%%%%%%%%%%%%%%%%%%%%%%%%%%%%%%%%%%%%%%%%%%%%%%%%%%%%%%%%%%%%%%%%%%%%%%%%%%%%%%%%%%%%%%%%%%%%%%%%%%%%%%%%%%%%%%%%%%%%%%%%%%%%%%%%%%%%%%%%%%%%%%%%%%%%%%%%%%%%%%%%%%%%%%%%%%%%%%%%%%%%%%%%%%
\section*{Declarations}
\subsection*{Funding}
This work was supported by Helmholtz PoF MT-DMA and used the Maxwell computational resources at DESY. 

\subsection*{Code Availability}
All code to generate and analyze the results presented in this manuscript is publicly available via \href{https://gitlab.desy.de/desy_dma/beamtimecompressionbenchmark}{GitLab}, including a \emph{Singularity}-container that was used for all experiments. Raw experiments are encapsulated into \texttt{Make}-targets for maximum reproducibility. 
\subsection*{Data Availability}
Intermediate results are available from the corresponding author upon reasonable request. The considered data corpus is not publicly available since it was partially generated by external PETRA III users. Upon reasonable request, the corresponding author will establish contact with the respective data owners and beamline scientists in charge.

%%%%%%%%%%%%%%%%%%%%%%%%%%%%%%%%%%%%%%%%%%%%%%%%%%%%%%%%%%%%%%%%%%%%%%%%%%%%%%%%%%%%%%%%%%%%%%%%%%%%%%%%%%%%%%%%%%%%%%%%%%%%%%%%%%%%%%%%%%%%%%%%%%%%%%%%%%%%%%%%%%%%%%%%%%%%%%%%%%%%%%%%%%%%%%%%%%%%%%%%%%%%%%%
\begin{appendices}
\onecolumn
\section{Compression Configurations}
\label{sec:compressors}
The following list of nine tables, Tab.~\ref{tab:brotli} to Tab.~\ref{tab:7z}, contains all configurations we considered in our analysis for each of the compressor tools.

\begin{table*}[b]%[htbp]
\centering
\caption{Brotli compression using the CLI tool \code{brotli} version 1.1.0 with options \code{--force} (overwrite output) and \code{--output} (output file), along with the parameters listed in this table.}
\label{tab:brotli}
\title{\textbf{Brotli}}
\begin{tabularx}{\textwidth}{l | l | c | l }
\toprule Parameter& Type & Values & Description \\\midrule
\multicolumn{4}{l}{\textbf{General:}} \\
\texttt{-q} & int & [0,11] & Compression level\\
\multicolumn{4}{l}{} \\
\multicolumn{4}{l}{\textbf{Window mode} (select one):} \\
\texttt{-w} & int & [10,24] & LZ77 window size $2^x-16$ bytes\\
\texttt{--large\_window} & int & [10,30] & Brotli large-window mode $2^x-16$ bytes\\
\end{tabularx}

\end{table*}

\begin{table*}%[htbp]
\centering
\caption{gzip compression using the CLI tool \code{gzip} version 1.12 with options \code{--force} (overwrite output), \code{-keep} (keep input), and \code{--stdout} (write to \code{stdout}), along with the parameters listed in this table.}
\title{\textbf{gzip}}
\label{tab:gzip}
\begin{tabularx}{\textwidth}{l | l | c | l }
\toprule Parameter& Type & Values & Description \\\midrule
\multicolumn{4}{l}{\textbf{General:}} \\
\texttt{-\#} & int & [1,9] & Compression level\\
\end{tabularx}
\end{table*}

\begin{table*}%[ht!]
\centering
\caption{lrzip compression using the CLI tool \code{lrzip} version 0.651 with options \code{--force} (overwrite output), \code{--outfile} (output file), and \code{--threads} (multi-threading), along with the parameters listed in this table.}
\label{tab:lrzip}
\title{\textbf{lrzip}}
\begin{tabularx}{\textwidth}{l | l | c | X }
\toprule Parameter& Type & Values & Description \\\midrule
\multicolumn{4}{l}{\textbf{General:}} \\
\texttt{--\#} & cat & \{lzma, lzo, bzip2, zpaq, gzip\} & Backend compressor\\
\texttt{-L} & int & [1,9] & Compression level\\
\texttt{-T} & flag & \{present, absent\} & Disable LZ4 compressibility testing\\
\multicolumn{4}{l}{} \\
\multicolumn{4}{l}{\textbf{Window mode} (select one):} \\
\texttt{-w} & int &\{1, 2, 4, 8, 16, 32, 64, 128\}& Window size in units of 100 MB\\
\texttt{-U} & flag & present & Unlimited window size\\
\end{tabularx}
\end{table*}

\begin{table*}%[ht!]
\centering
\caption{lzop compression using the CLI tool \code{lzop} version 1.04 with options \code{-f} (overwrite output) and \code{-o} (output file), along with the parameters listed in this table.}
\label{tab:lzop}
\title{\textbf{lzop}}
\begin{tabularx}{\textwidth}{l | l | c | X }
\toprule Parameter& Type & Values & Description \\\midrule
\multicolumn{4}{l}{\textbf{General:}} \\
\texttt{-\#} & int & [1,9] & Compression level\\
\end{tabularx}
\end{table*}

\begin{table*}%[ht!]
\centering
\caption{LZ4 compression using the CLI tool \code{lz4} version 1.10.0 with options \code{-T\#} (multi-threading), along with the parameters listed in this table.}
\label{tab:lz4}
\title{\textbf{LZ4}}
\begin{tabularx}{\textwidth}{l | l | c | X }
\toprule Parameter& Type & Values & Description \\\midrule
\multicolumn{4}{l}{\textbf{General:}} \\
\texttt{-B\#} & int & [4,7] & Block size\\
\texttt{-BI} / \texttt{-BD} & flag & \{on, off\} & Block independence\\
\texttt{--favor-decSpeed} & flag & \{present, absent\} & Optimize for decompression speed\\
\multicolumn{4}{l}{} \\
\multicolumn{4}{l}{\textbf{Compression mode} (select one):} \\
\texttt{-\#} & int & [1,12] & Normal mode\\
\texttt{--fast} & int & \multicolumn{1}{l|}{\{1, 3, 5, 7, 10, 15,} & Fast mode\\
 &  & \multicolumn{1}{r|}{20, 30, 50, 100, 150\}} & \\
\end{tabularx}
\end{table*}

\begin{table*}%[ht!]
\centering
\caption{XZ compression using the CLI tool \code{xz} version 5.6.3 with options \code{--stdout} (write to \code{stdout}), \code{-keep} (keep input), and \code{--threads} (multi-threading), along with the parameters listed in this table. We use either the preset mode or the custom filter.}
\label{tab:xz}
\title{\textbf{XZ Utils}}
\begin{tabularx}{\textwidth}{l | l | c | X }
\toprule Parameter& Type & Values & Description \\\midrule
\multicolumn{4}{l}{\textbf{Preset mode:}} \\
\texttt{-\#} & int & [0,9] & Compression level\\
\texttt{--extreme} & flag & \{present, absent\} & Extreme mode\\
\texttt{--block-size} & int & \{$2^i$MiB: $i\in[0,9]$\} & Block size\\
\multicolumn{4}{l}{} \\
\multicolumn{4}{l}{\textbf{Custom filter mode:}} \\
\texttt{--lzma2} & flag & present & Custom compressor filter\\
\texttt{dict} & str & \{$2^i$MiB: $i\in[0,11]$\} & Dictionary size\\
\texttt{mf} & cat & \{hc3, hc4, bt2, bt3, bt4\} & Match finder\\
\texttt{mode} & cat & \{fast, normal\} & Compression mode\\
\texttt{depth} & int &  $\{0\}\cup \{2^i : i\in[2,10]\}$ & Maximum search depth\\
\texttt{nice} & int & \{8, 16, 32, 64, 96, 192, 273\} & Target match length in bytes\\
\end{tabularx}
\end{table*}

\begin{table*}%[ht!]
\centering
\caption{ZPAQ compression using the CLI tool \code{zpaq} version 7.15 with options \code{-threads} (multi-threading), and \code{-force} (overwrite output), along with the parameters listed in this table. We use the general options together with either the preset mode, BWT mode, or the LZ77 mode.}
\label{tab:zpaq}
\title{\textbf{ZPAQ}}
\begin{tabularx}{\textwidth}{l | l | c | X }
\toprule Parameter& Type & Values & Description \\\midrule
\multicolumn{4}{l}{\textbf{General:}} \\
\texttt{-fragment} & int & [4,12] & Fragment size range\\
\texttt{Blocksize} & int & absent or [0,11] & Block size\\
\multicolumn{4}{l}{} \\
\multicolumn{4}{l}{\textbf{Preset Mode:}} \\
\texttt{-m\#} & int & [0,5] & Compression level\\
\multicolumn{4}{l}{} \\
\multicolumn{4}{l}{\textbf{BWT mode:}} \\
\texttt{pre} & int & \{3\} & Preprocessing model\\
\texttt{comp} & cat & \multicolumn{1}{l|}{\{ci1, ci1m, ci1ms, ci1mst,} & Context model settings\\
&  & \multicolumn{1}{r|}{ci1w, ci1mw, ci1msw\}} & \\
\multicolumn{4}{l}{} \\
\multicolumn{4}{l}{\textbf{LZ77 mode:}} \\
\texttt{pre} & int & \{1, 2\} & Preprocessing model\\
\texttt{comp} & cat & \multicolumn{1}{l|}{\{ci1, ci1m, ci1ms, ci1mst,} & Context model settings\\
&  & \multicolumn{1}{r|}{ci1w, ci1mw, ci1msw\}} & \\
\texttt{min1} & int & \{1, 3\} & Minimum match length\\
\texttt{min2} & int & \{0, 4\} & Longer minimum match length\\
\texttt{depth} & int & \{4, 6\} & Search depth\\
\texttt{size} & int & \{24, 26\} & Hash table size\\
\texttt{lookahead} & int & \{0, 1\} & Match lookahead\\
\end{tabularx}
\end{table*}

\begin{table*}%[ht!]
\centering
\caption{Zstandard compression using the CLI tool \code{zstd} version 1.5.7 with options \code{-o} (output file), \code{-T\#} (multi-threading), and \code{--force} (overwrite output), along with the parameters listed in this table.}
\label{tab:zstd}
\title{\textbf{Zstandard}}
\begin{tabularx}{\textwidth}{l | l | c | X }
\toprule Parameter& Type & Values & Description \\\midrule
\multicolumn{4}{l}{\textbf{General:}} \\
\texttt{--long} & int & absent or [10,31] & Long distance matching window\\
\texttt{--[no-]compress-literals} & flag & \{on, off\} & Literal compression\\
\texttt{--[no-]row-match-finder} & flag & \{on, off\} & Enable/disable row-based match finder\\
\multicolumn{4}{l}{} \\
\multicolumn{4}{l}{\textbf{Compression mode} (select one):} \\
\texttt{-\#} & int & [0,19] & Normal mode\\
\texttt{--ultra} & int & [20,22] & Ultra mode\\
\texttt{--fast} & int & \multicolumn{1}{l|}{\{1, 2, 3, 4, 5, 7, 10, 15,} & Fast mode\\
 &  & \multicolumn{1}{r|}{20, 30, 50, 100, 150\}} & \\
\end{tabularx}
\end{table*}

\begin{table*}%[ht!]
\centering
\caption{7-Zip compression using the CLI tool \code{7z} version 23.01 with options \code{-mmt} (multi-threading), along with the parameters listed in this table. We use the general options together with one of the listed modes.}
\label{tab:7z}
\title{\textbf{7-Zip}}
\begin{tabularx}{\textwidth}{l | l | c | X }
\toprule Parameter& Type & Values & Description \\\midrule
\multicolumn{4}{l}{\textbf{General:}} \\
\texttt{-t\#} & cat & \{7z, zip\} & Type of archive\\

\multicolumn{4}{l}{} \\
\multicolumn{4}{l}{\textbf{Preset mode:}} \\
\texttt{-mx} & int & [1,9] & Compression level\\

\multicolumn{4}{l}{} \\
\multicolumn{4}{l}{\textbf{bzip2 or DEFLATE mode:}} \\
\texttt{-m0} & cat & \{bzip2, deflate\} & Compression algorithm\\
\texttt{-mx} & int & [1,9] & Compression level\\

\multicolumn{4}{l}{} \\
\multicolumn{4}{l}{\textbf{LZMA or LZMA2 mode:}} \\
\texttt{-m0} & cat & \{lzma, lzma2\} & Compression algorithm\\
\texttt{-md} & str & \multicolumn{1}{l|}{\{1m, 2m, 4m, 8m, 16m,} & Dictionary size\\
 &  & \multicolumn{1}{r|}{32m, 128m, 256m, 512m, 1g\}} & \\
 \texttt{-mfb} & int & \{32, 64, 128, 273\} & Number of fast bytes\\
 \texttt{-mmf} & cat & \{bt2, bt3, bt4, hc4\} & Match finder\\

\multicolumn{4}{l}{} \\
\multicolumn{4}{l}{\textbf{PPMd mode:}} \\
\texttt{-m0} & cat & \{ppmd\} & Compression algorithm\\
\texttt{-mo} & int & [6, 32] & Model order\\
\texttt{-mmem} & int & [18, 42] & Memory size\\

\end{tabularx}
\end{table*}
\end{appendices}

\clearpage
\bibliography{References}

\end{document}